\documentclass[aps,prb,twocolumn,groupaddress,longbibliography]{revtex4-1}
\usepackage{epsfig,amsopn}
\usepackage{graphicx}
\usepackage{color}
\usepackage{amsmath,amssymb}
\usepackage{enumerate}
\newcommand\bea{\begin{eqnarray}}
\newcommand\eea{\end{eqnarray}}
\newcommand\beq{\begin{equation}}
\newcommand\eeq{\end{equation}}

\def\nn{\nonumber}
\def\f{\frac}
\def\al{\alpha}

\def\ep{\epsilon}

\def\si{\sigma}
\def\Do{\partial}

\def\dg{\dagger}
\def\la{\langle}
\def\ra{\rangle}
\def\ua{\uparrow}
\def\da{\downarrow}

\def\ka{\kappa}
\def\th{\theta}

\begin{document}
\title{Transverse currents in spin transistors} 
\author{Bijay Kumar Sahoo}
\affiliation{School of Physics, University of Hyderabad, Prof. C. R. Rao Road, Gachibowli, Hyderabad-500046, India}
\author{ Abhiram Soori~~}
\email{abhirams@uohyd.ac.in}
\affiliation{School of Physics, University of Hyderabad, Prof. C. R. Rao Road, Gachibowli, Hyderabad-500046, India}
\begin{abstract}
In many systems, planar Hall effect wherein transverse signal appears in response to longitudinal stimulus is rooted in spin-orbit coupling. A spin transistor put forward by Datta and Das on the other hand consists of ferromagnetic leads connected to spin-orbit coupled central region and its conductance can be controlled by tuning the strength of spin-orbit coupling. We find that transverse currents also appear in Datta-Das transistors made by connecting two  two-dimensional ferromagnetic reservoirs to  a central spin-orbit coupled two-dimensional electron gas. We find that the spin transistor exhibits a nonzero transverse conductivity which  depends on the direction of polarization in ferromagnets and the location where it is measured. We study the conductivities for the system with finite and infinite widths. The conductivities exhibit Fabry-P\'erot type oscillations as the length of the spin-orbit coupled regions is varied. Interestingly, even in the limit when longitudinal conductivity is made zero by cutting off the junction between the central spin-orbit coupled region and the ferromagnetic lead on one side (right), the transverse conductivities remain nonzero in the regions that are on the left side of the cut-off junction.  
\end{abstract}
\maketitle
\section{Introduction}
In a two-dimensional metal, a transverse voltage results in response to a current in presence of a magnetic field normal to the plane of the metal, an effect known as Hall effect~\cite{kittel}. This is due to Lorentz force on electrons in the metal. It was found that in certain systems, the voltage developed perpendicular to the current, magnetic field and the current - all three can lie in the same plane, an effect known as planar Hall effect~\cite{goldberg54,tang03,Roy10,annadi13,taskin17,he19,bharadwaj21,burkov17,kumar18,sonika21}. While in many of these systems~\cite{goldberg54,tang03,annadi13}, the planar Hall effect is due to spin-orbit coupling (SOC), the origin is rooted in chiral anomaly in some other systems~\cite{burkov17,kumar18}. In topological insulators, spin momentum locking - which is  qualitatively same as SOC, causes this effect~\cite{suri21}. In spin-orbit coupled  metals, the transverse deflection of the longitudinal current  under the influence of Zeeman field explains  planar Hall effect~\cite{soori2021}. A closely related phenomenon -  ``in-plane Hall effect" rooted in Berry curvature and band geometric effects has also been reported recently~\cite{Liang2018,Zhou2022,wang22}.

Datta-Das transistor proposed in 1990 makes use of the fact that the  electron spin precesses in spin-orbit coupled region~\cite{dattadas}. Experimentally, it was challenging to realize such a transistor, since good quality systems with Rashba spin split bands and spin polarized electrons in semiconductors were difficult to achieve. It was demonstrated in 1997 that the strength of SOC in a semiconductor can be tuned by an applied gate voltage~\cite{nitta1997}. The first version of spin transistor proposed by Datta and Das was realized in 2009~\cite{koo09}. Around the same time, transverse signal was detected in spin-orbit coupled systems by injecting spin polarized electrons~\cite{Wunderlich2009}. Later, improved versions of Datta-Das transistor were developed~\cite{chuang2015,Choi2015}. In the improved versions~\cite{chuang2015,Choi2015}, the transport was ballistic, in contrast to the diffusive transport in the earlier versions~\cite{koo09,Wunderlich2009}. Transport in Datta-Das transistor has also been investigated theoretically~\cite{aharony2019,sarkar2020}. 

 In `planar Hall effect', the transverse voltage is due to deflection of longitudinal current in the spin-orbit coupled region by the application of in-plane Zeeman field. 
 It would be interesting to investigate whether there is a transverse deflection in the spin-orbit coupled region when the injected electrons are spin-polarized, instead of using a Zeeman field. In this work, we explore the possibility of a transverse current in response to a bias in a   ferromagnet-spin-orbit coupled region-ferromagnet junction, where the two ferromagnets are parallel. We follow Landauer-B\"uttiker scattering approach generalized to two-dimensional systems to address this problem~\cite{landauer1957r,buttiker1985m,datta1995}.  We find that the value of transverse conductivity depends on the location and the spin polarization of the ferromagnets. We study infinitely wide systems as well as the ones having finite width. Further, we find that the transverse conductivity is nonzero even in the case when the one of the junctions between the spin-orbit coupled region and the ferromagnet is cut-off and the longitudinal conductivity is zero.

 \begin{figure}
  \includegraphics[width=8cm]{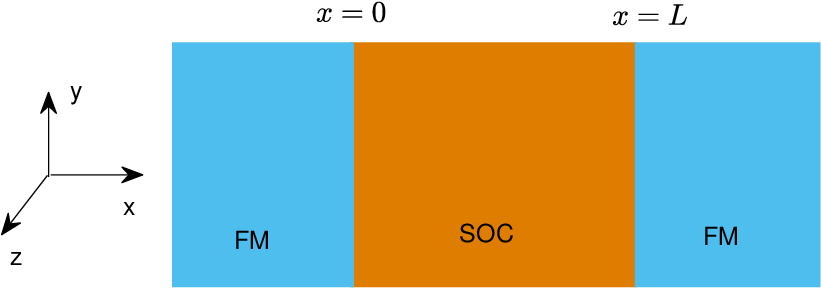}
  \caption{Schematic diagram of the setup. A spin-orbit coupled two-dimensional electron gas is connected to ferromagnets on either sides. }\label{fig:schem}
 \end{figure}

\section{Details of calculation}~\label{sec:calc}
The Hamiltonian describing the setup depicted in Fig.~\ref{fig:schem} is
\bea H &=& \Big[-\f{\hbar^2}{2m}\Big(\f{\Do^2}{\Do x^2}+\f{\Do^2}{\Do y^2}\Big)-\mu\Big]\si_0 + b\si_{\th,\phi}, \nn \\ 
&& {\rm for ~~}x<0, ~~{\rm and }~~x>L,  \nn \\
&=& \Big[-\f{\hbar^2}{2m}\Big(\f{\Do^2}{\Do x^2}+\f{\Do^2}{\Do y^2}\Big)-\mu\Big]\si_0 + i\al\Big(\si_y\f{\Do}{\Do x}-\si_x\f{\Do}{\Do y}\Big), \nn \\ 
&& {\rm for ~~}0<x<L,\label{eq:ham}
\eea
where $\si_{\th,\phi}=\cos{\th}\si_z+\sin{\th}(\cos{\phi}\si_x+\sin{\phi}\si_y)$. Here, $m$ is the effective mass of electrons in the system, $\mu$ -the chemical potential, $\al$ -the strength of SOC, $b$ -the magnitude of the Zeeman energy that characterizes the ferromagnet,  $(\th,\phi)$ -the direction of spin polarization of the ferromagnets and $\si_0, \si_x,\si_y, \si_z$ are Pauli spin matrices. The region $0<x<L$ is spin-orbit coupled region. The regions $x<0$ and $x>L$ are  ferromagnetic wherein the electrons are spin polarized. Along $y$-direction the system is assumed to be translationally invariant.  Current is conserved at the junctions for  very general boundary conditions other than continuity of wavefunction and the derivatives, akin to the problem of  point like scatterer in one dimension~\cite{soori20}. Here we choose the wavefunction and the derivatives  to satisfy 
\bea 
\psi(x_0^{-s_0}) &=& c(x_0) \psi(x_0^{s_0}) \nn \\ 
c(x_0) \Do_x\psi|_{x_0^{-s_0}} &=& \Do_x\psi|_{x_0^{s_0}} -\f{i\al m}{\hbar^2} \si_y\psi|_{x_0^{s_0}}, \label{eq:BC} \eea
at $x_0=0,L$ where $s_0={\rm sign}(L/2-x_0)$ and $x_0^{s_0}=lim_{\ep\to 0^+}[x_0+s_0\ep]$. Here, $c(x_0)$ is a real constant. The value $c(x_0)=0$ implies that the junction is cut off at $x_0$.  In the ferromagnetic leads ($x<0$ and $x>L$), the dispersion relations are $E=E_{\pm}=\hbar^2(k_x^2+k_y^2)/2m-\mu\pm b$, and  if $b>\mu>0$, near zero energy, only one band $E=E_-$ exists. We shall consider transport in the energy range $-b-\mu<E<b-\mu$ where only one band actively participates in the transport.  In this energy range, the spins of the plane wave modes in the ferromagnet are polarized in a direction $(\pi-\th,-\phi)$. In the spin-orbit coupled region ($0<x<L$), the dispersion is $E=\hbar^2k^2/2m-\mu\pm\al k$, where $k=\sqrt{k_x^2+k_y^2}$. The width $W$ in the $y$-direction can be either finite or infinite. For a finite $W$ with periodic boundary conditions in $y$-direction, the $k_y$ is quantized whereas in the limit of $W\to\infty$, $k_y$ takes continuous values and the electron can be incident at any angle $-\pi/2<\chi<\pi/2$. 

\subsection{$W\to\infty$}
The wavefunction of an electron that approaches from the left ferromagnet onto the SOC region at energy $E$ making an angle $\chi$ with $x$-axis has the form $\psi(x)e^{ik_yy}$ (where $k_y=k\sin{\chi}$ and $k=\sqrt{2m(E+\mu+b)}/\hbar$) with 
\bea 
\psi(x) &=& (e^{ik_xx} + r_ke^{-ik_xx})|\da\ra + r'_ke^{\ka x}|\ua\ra ~~{\rm for ~~} x\le 0, \nn \\ &=& \sum_{j=1}^4 s_j e^{ik_{x,j}x} [u_{j},~v_j]^T ~~{\rm for ~~} 0\le x \le L, \nn \\ 
&=& t_ke^{ik_xx}|\da\ra +t'_ke^{-\ka (x-L)}|\ua\ra ~~{\rm for~~}x\ge L, \label{eq:psi}
\eea
where ${|\ua\ra}$ and ${|\da\ra}$ are eigenspinors of $\si_{\th,\phi}$ with ${|\ua\ra}=[\cos{(\th/2)},~ \sin{(\th/2)}e^{i\phi}]^T$, ${|\da\ra}=[-\sin{(\th/2)},~\cos{(\th/2)}e^{i\phi}]^T$, $k_{x,j}$ for $j=1,2,3,4$ are given by 
$k_{x,1}=\sqrt{k_{+}^{'2}-k_y^2}$,  $k_{x,2}=-\sqrt{k_{+}^{'2}-k_y^2}$, $k_{x,3}=\sqrt{k_{-}^{'2}-k_y^2}$, and  $k_{x,4}=-\sqrt{k_{-}^{'2}-k_y^2}$, with $k'_{\si}= \sqrt{2[\al^2+\hbar^2(E+\mu)/m+\si\al\sqrt{\al^2+2\hbar^2(E+\mu)/m}]}~m/\hbar^2$ for  $\si=+,-$, and $u_j=\al{(k_y+ik_{x,j})}$, $v_j=(E+\mu)-\hbar^2(k_{x,j}^2+k_y^2)/2m$.  The boundary conditions at $x=0, L$, can be utilized to determine the scattering amplitudes- $r_k, r'_k, t_k, t'_k, s_j$. The differential conductivity $G_{xx}$ which is defined as $dI_x/dV$ where $dI_x$ and $dV$ are  infinitesimal changes in current density along $\hat x$-direction  $I_x$ and the bias $V$ respectively, is given by  
\bea 
G_{xx} &=& \f{e^2}{h} \f{k}{2\pi} \int_{-\pi/2}^{\pi/2}d\chi \cos{\chi}~|t_k|^2. 
\label{eq:Gx} \eea
Here, $k$ is evaluated at energy $E=eV$. 

The net current in the $y$-direction $I_y$ can also have a nonzero value. Transverse conductivity $G_{yx}$ defined as $dI_y/dV$ where $dI_y$ is  the infinitesimal change in  current density along $\hat y$-direction  is given by 
\bea
G_{yx}(x) &=& \f{e}{h} \f{m}{h} \int_{-\pi/2}^{\pi/2}d\chi I_y(\chi,x), \label{eq:Gy} 
\eea
where $I_{y}(\chi,x)$ is the current density along $y$-direction for angle of incidence $\chi$ at location $x$. The current density along $y$-direction $I_{y}(\chi,x)$ depends on $x$ - the location  in the longitudinal direction and angle of incidence $\chi$. It is given by 
\bea 
I_y(\chi,x) &=& \f{e\hbar k_y}{m}\psi^{\dag}(x)\psi(x), ~~{\rm for ~~}x<0~{\rm and~~}x>L \nn \\ 
&=& e\Big[ \f{\hbar k_y}{m}\psi^{\dag}(x)\psi(x)+\f{\al}{\hbar} \psi^{\dag}(x)\si_x\psi(x)\Big], \nn \\ &&~~~ ~~{\rm for ~}0<x<L. \nn \\ &&  \label{eq:Iy}
\eea

\begin{figure*}
 \includegraphics[width=5.5cm]{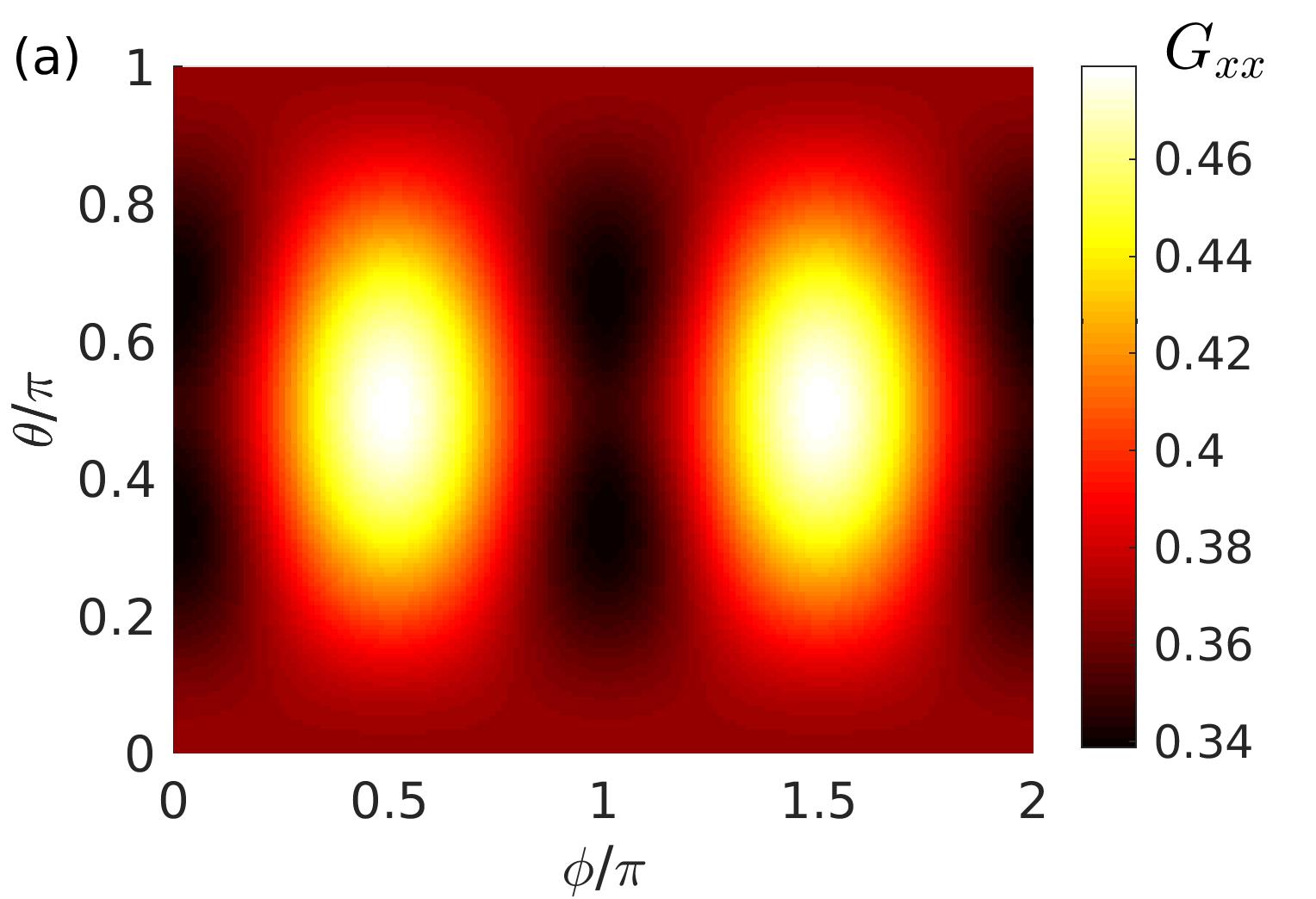}
 \includegraphics[width=5.5cm]{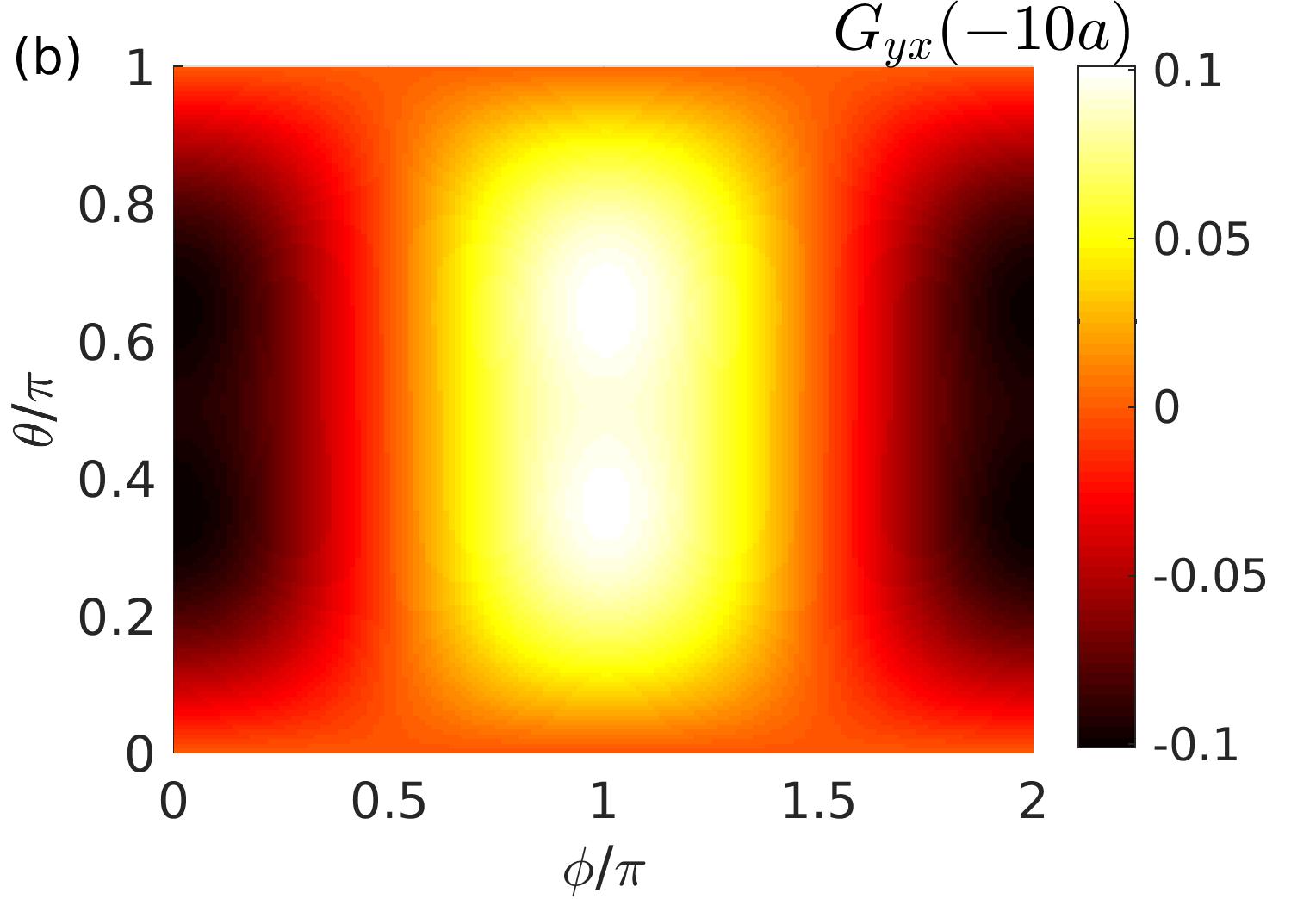}
 \includegraphics[width=5.5cm]{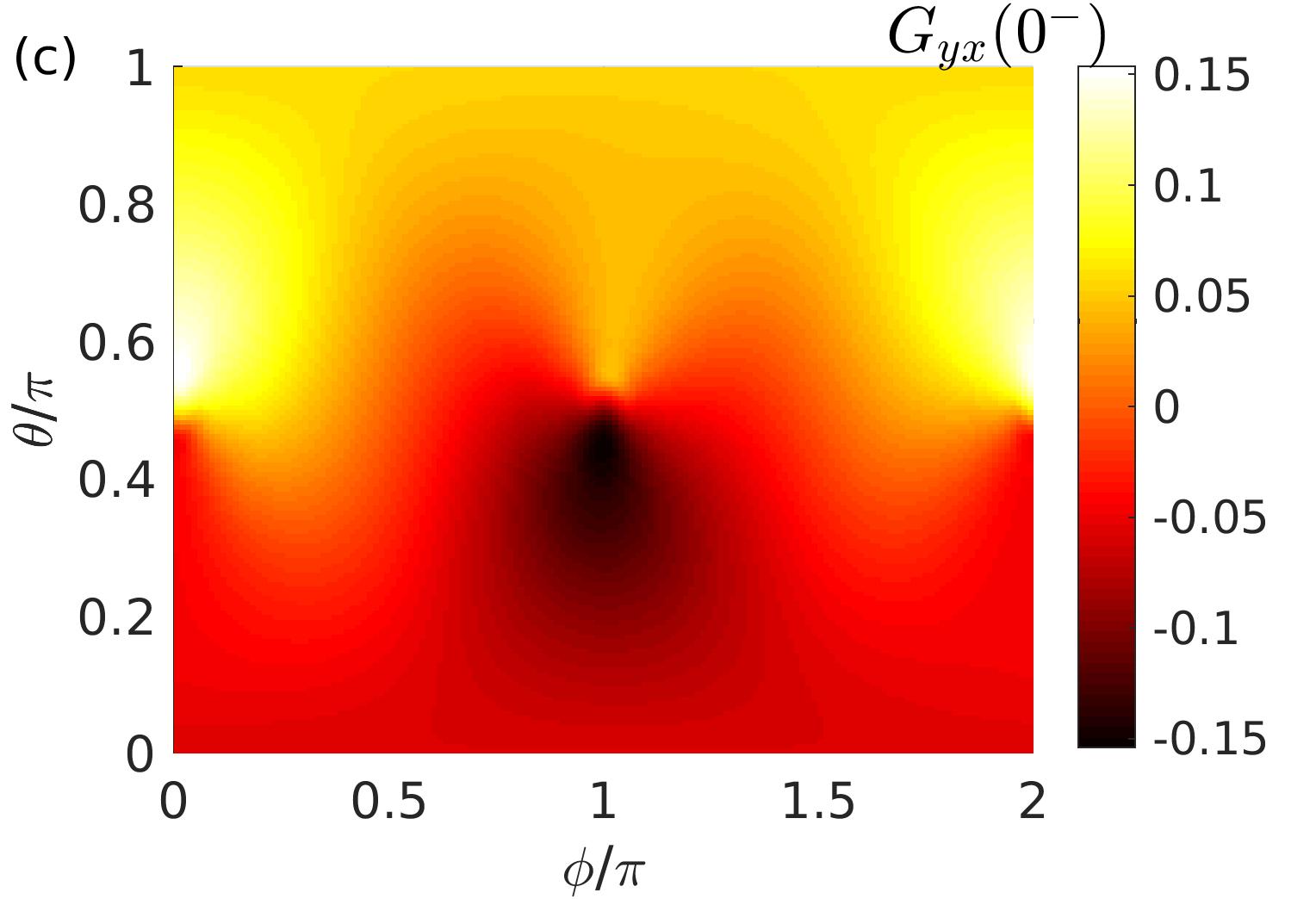}
 \includegraphics[width=5.5cm]{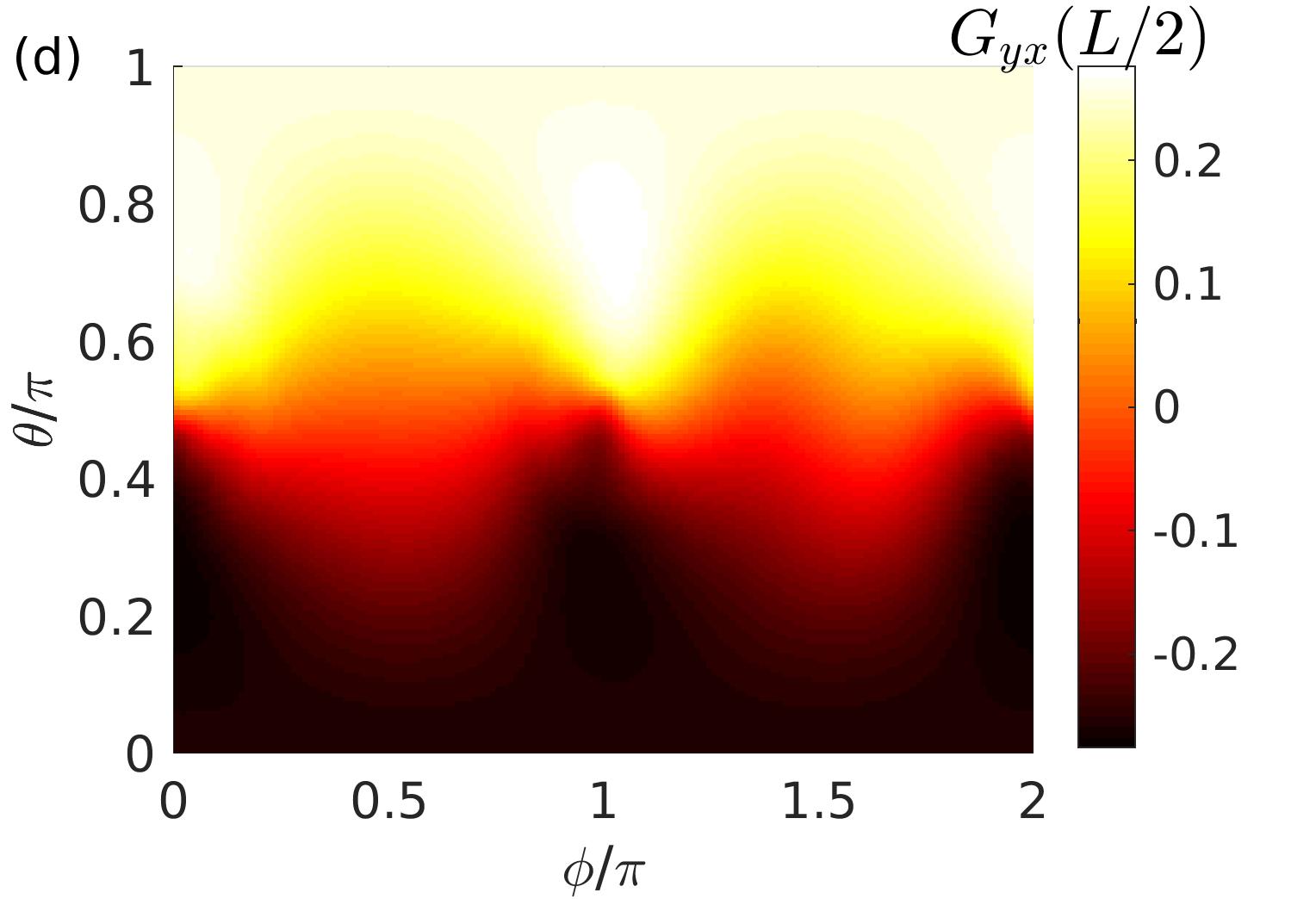}
 \includegraphics[width=5.5cm]{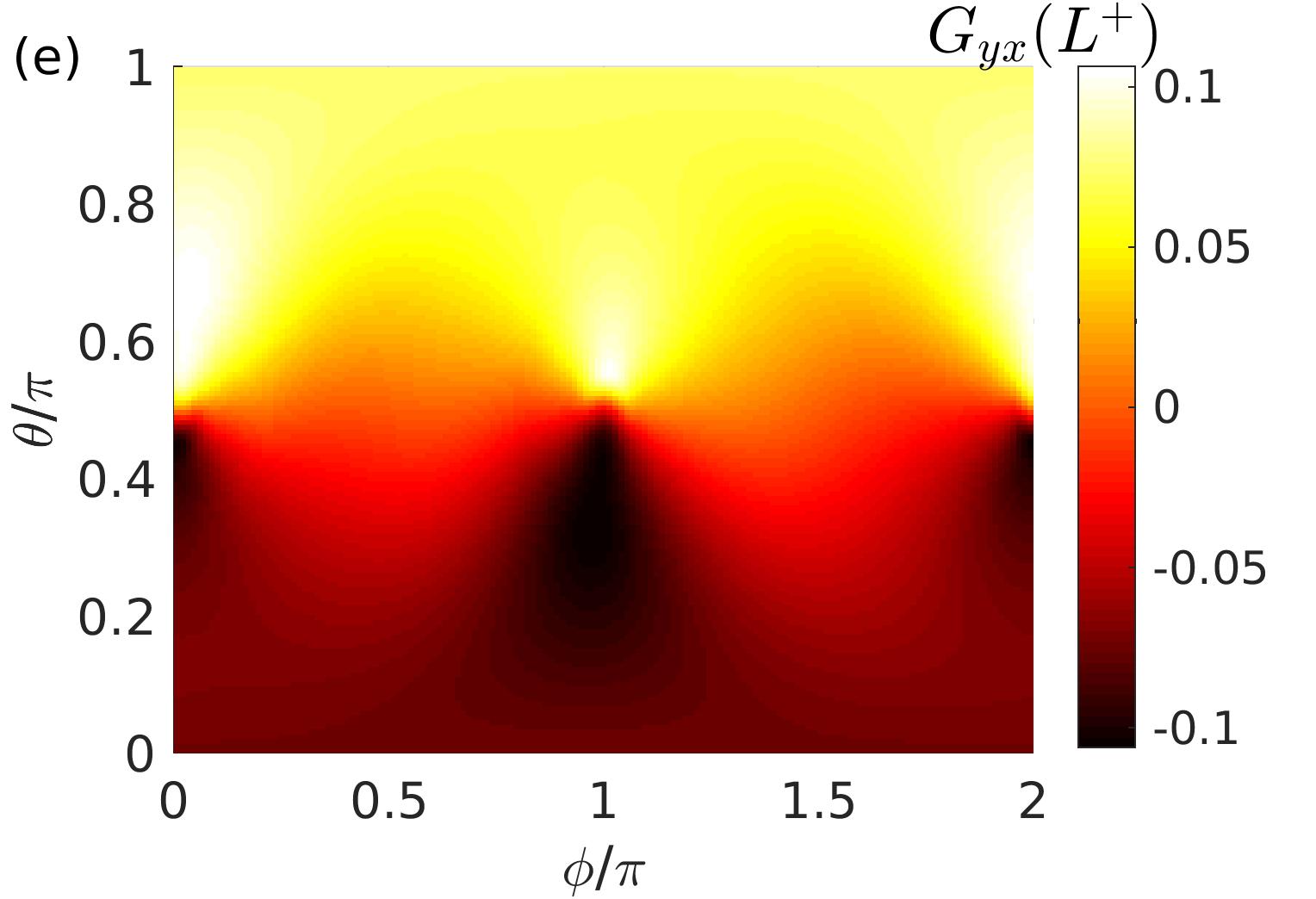}
 \includegraphics[width=5.5cm]{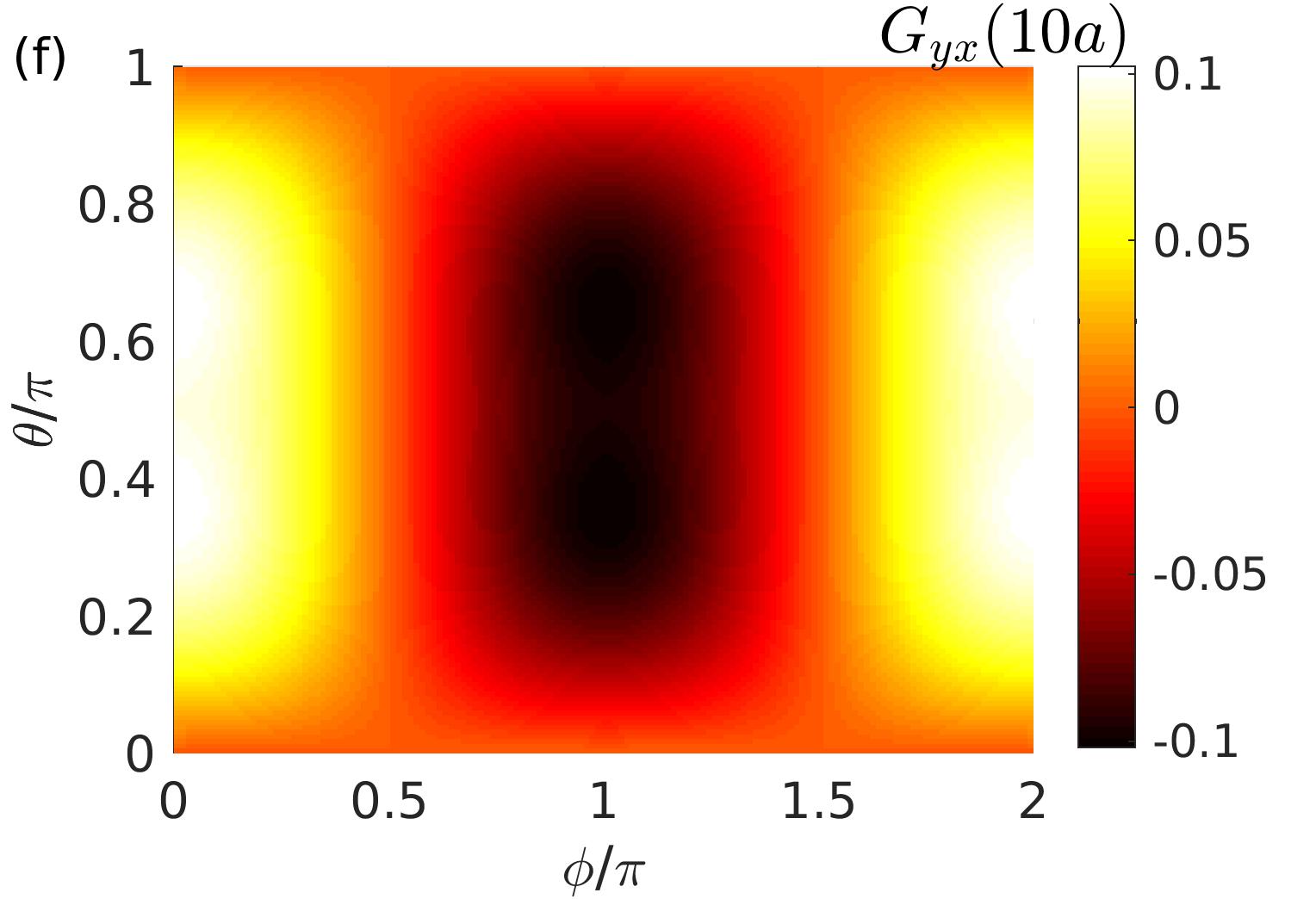}
 \caption{(a) Differential conductivity $G_{xx}$ versus $\th$ and $\phi$. Differential transverse conductivity $G_{yx}$ is plotted versus $\th$ and $\phi$ at different locations $x$: (b) $x=-10a$, (c)$x=0^-$, (d) $x=L/2$, (e) $x=L^+$, (f) $x=L+10a$.  All conductivities are in units of $e^2/ha$. Parameters: $\al=0.5\hbar\sqrt{\mu/m}$, $b=2\mu$, $L=5a$ (where $a=\hbar/\sqrt{m\mu}$) and $E=eV=0$.   }\label{fig:G}
\end{figure*}

\subsection{Finite $W$}
For finite width $W$ with periodic boundary conditions, $k_y$ takes values  that are integer multiples of $2\pi/W$. At energy $E$, the modes that contribute to transport are the ones with $k_y=n2\pi/W$ where $n$ takes integer values in the range $-N_y\le n\le N_y$, $N_y=[kW/(2\pi)]$ ($[x]$ denotes the maximum integer less than or equal to $x$), $k=\sqrt{2m(E+\mu+b)}/\hbar$. The wavefunction for an electron incident in a mode with a given $k_y$ takes the form $\psi_n(x)e^{ik_yy}$ with
\bea 
\psi_n(x) &=& (e^{ik_xx} + r_{k,n}e^{-ik_xx})|\da\ra + r'_{k,n}e^{\ka x}|\ua\ra ~~{\rm for ~~} x\le 0, \nn \\ &=& \sum_{j=1}^4 s_{j,n} e^{ik_{x,j}x} [u_{j},~v_j]^T ~~{\rm for ~~} 0\le x \le L, \nn \\ 
&=& t_{k,n}e^{ik_xx}|\da\ra +t'_{k,n}e^{-\ka (x-L)}|\ua\ra ~~{\rm for~~}x\ge L, \label{eq:psi-W}
\eea
where $k_x=\sqrt{k^2-k_y^2}$. The notation is similar to the one followed next to eq.~\eqref{eq:psi}, except that the scattering coefficients depend on the index $n$ that dictates the value of $k_y$. 
The differential conductivity $G_{xx}$ is given by the expression 
\bea  
G_{xx} &=& \f{e^2}{h} \f{1}{W}\sum_{n=-N_y}^{N_y} |t_{k,n}|^2~.  \label{eq:Gxx-W}
\eea
The transverse conductivity is given by the expression
\bea
G_{yx}(x) &=& \f{e}{h}\f{m}{\hbar W}\sum_{n=-N_y}^{N_y} \f{I_{y,n}}{k_x} \label{eq:Gyx-W}, 
\eea
where 
\bea
I_{y,n} &=& \f{e\hbar}{m}k_y \psi^{\dag}_n(x)\psi_n(x), ~~{\rm for ~~}x<0 ~~{\rm and ~}x>L, \nn \\ 
&=&  e\big[ \f{\hbar k_y}{m}\psi_n^{\dag}(x)\psi_n(x)+\f{\al}{\hbar} \psi_n^{\dag}(x)\si_x\psi_n(x) \big] \nn \\ &&~~~~~~~~~{\rm for ~}0<x<L.
\eea
Note that here $k_x, k_y$ depend on $n$.

\section{Results and Analysis}\label{sec:res}

\begin{figure}[htb]
 \includegraphics[width=8cm]{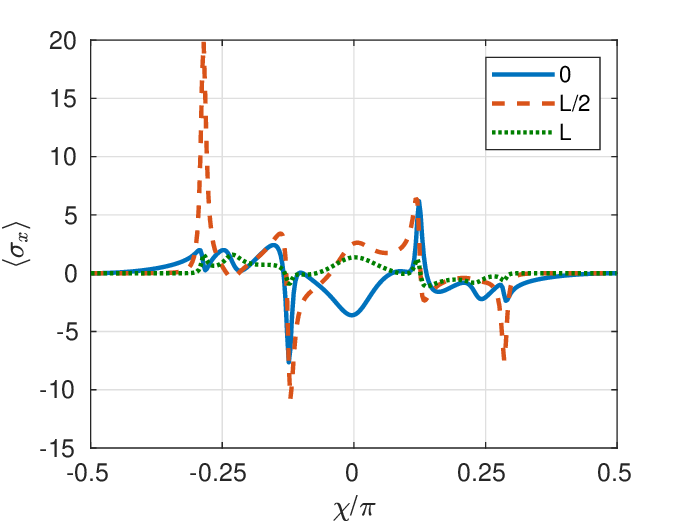}
 \caption{$\la\si_x\ra$ evaluated using the wavefunction $\psi(x)$ at $x=0^-, L/2, L^+$ for $\th=\pi/4$ and $\phi=\pi/2$ for  $\al=0.5\hbar\sqrt{\mu/m}$, $b=2\mu$, $L=5a$ (where $a=\hbar/\sqrt{m\mu}$) and $E=eV=0$. }\label{fig:sixchi}
\end{figure}

We choose  $\al=0.5\hbar\sqrt{\mu/m}$, $b=2\mu$, $L=5a$ (where $a=\hbar/\sqrt{m\mu}$), $c(0)=c(L)=1$, $E=eV=0$ and calculate the conductivity $G_{xx}$ and the transverse conductivity $G_{yx}(x)$ at $x=-10a, 0, L/2, L, L+10a$ numerically as functions of $\th$ and $\phi$ and plot in Fig.~\ref{fig:G}. Here zero bias conductivity means the ratio of tiny current density $dI$ as a response to a tiny bias $dV$ in the limit $dV\to 0$. The conductivity $G_{xx}$ depends on the spin polarization direction of the ferromagnet [see Fig.~\ref{fig:G}(a)] - a characteristic feature of Datta-Das spin transistor. When the spin polarization in the leads is along $\hat y$-direction, the incident electron spin is the same as the spin of the electrons moving along $\hat x$ direction in the spin-orbit coupled region, which contributes the most to the longitudinal conductivity. Hence, $G_{xx}$ is peaked around $(\pi/2,\pi/2)$ and $(\pi/2,3\pi/2)$ in Fig.~\ref{fig:G}(a). In Fig.~\ref{fig:G}(b-f), the transverse conductivities at different locations $x$ are plotted as functions of $(\th,\phi)$. The value of transverse conductivity depends on the location $x$  and the polarization direction. The dependence on the polarization direction is because of the SOC in the central region. 
The SOC term in the Hamiltonian is $\al(\si_xk_y-\si_yk_x)$. Any spin component $\la\si_x\ra$ will imply that the transverse wave numbers  $k_y$ and $-k_y$ are inequivalent and the scattering probabilities are different for these transverse wave numbers. The transverse conductivity is integral of $I_y$ over $\chi$. $I_y$ depends on $\chi$ through the wavefunction as shown in eq.~\eqref{eq:Iy} and if the scattering amplitudes change when $\chi$ goes to $-\chi$,  $G_{yx}$ acquires a nonzero value. 

\begin{figure}[htb]
 \includegraphics[width=8cm]{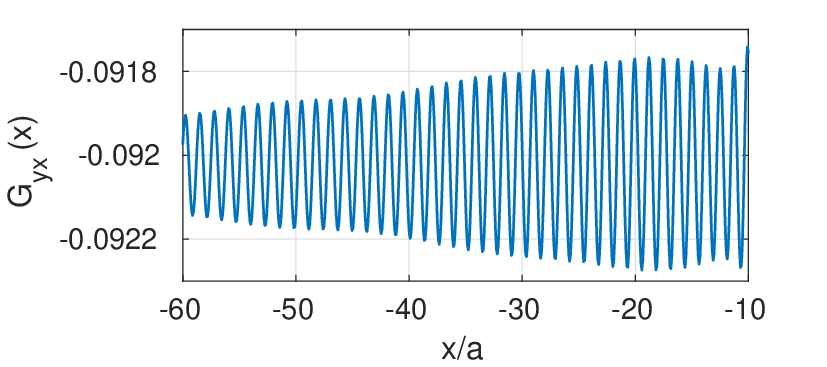}
 \caption{ Transverse conductivity as a function of the location in the region $x<0$ for $\th=\pi/4$ and $\phi=0$. Other parameters are the same as in Fig.~\ref{fig:G}. }\label{fig:GyxLL}
\end{figure}

A particularly interesting case is for $\phi=\pi/2,3\pi/2$ for which, the ferromagnets are polarized with  a zero value for $\la\si_x\ra$. This means that for $\phi=\pi/2,3\pi/2$, the transverse conductivity should be zero. While this holds true at $x=-10a,L+10a$, the transverse conductivities at locations $x=0, L/2, L$ are nonzero  [see  Fig.\ref{fig:G}(b-f)]. The reason for this feature is in the wavefunction which also has evanescent modes pointing along $|\ua\ra$ that decay away from the spin-orbit coupled region, but are present near the junctions at $x=0, L$ as can be seen from eq.~\eqref{eq:psi}. We plot $\la\si_x\ra$ using the wavefunction $\psi(x)$ at locations $x=0^-,L/2,L^+$ as a function of angle of incidence $\chi$ for $\th=\pi/4$ and $\phi=\pi/2$ in Fig.~\ref{fig:sixchi}. 
We find that not only $\la\si_x\ra$ is nonzero at $x=0, L/2, L$, it is not an odd function of $\chi$ and when $I_y$ is integrated over $\chi$, the transverse conductivity obtained is not zero for  $\phi=\pi/2$. 
The transverse conductivity in the regions $x\ll 0$ and $x\gg L$ is not affected by the interference of the evanescent modes with the plane waves and meets the expectation that when $\phi=\pi/2, 3\pi/2$,  $G_{yx}=0$. 

\begin{figure}[htb]
 \includegraphics[width=8cm]{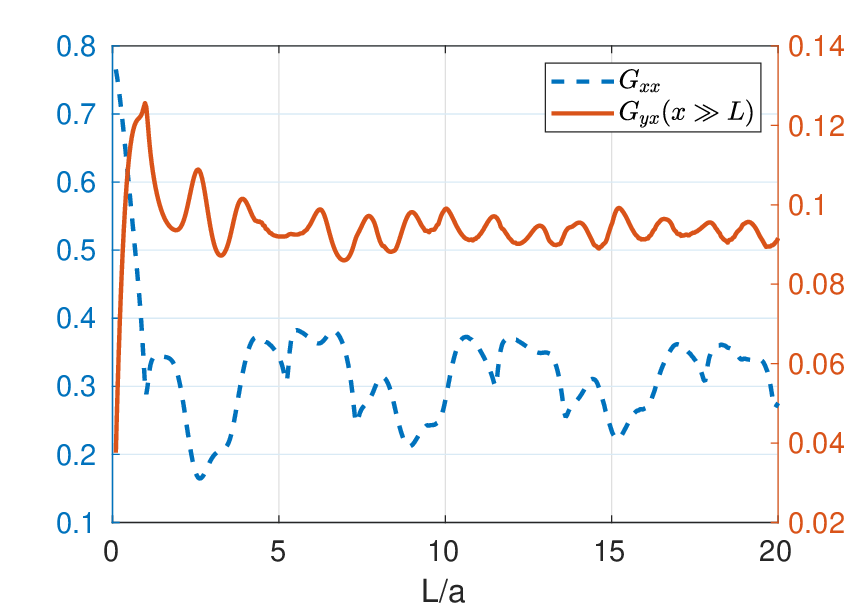}
 \caption{Conductivities in units of $e^2/ha$ versus $L$ for $\th=\pi/4$ and $\phi=0$. Other parameters are the same as in Fig.~\ref{fig:G}. }\label{fig:GvsL}
\end{figure}

Another feature of the transverse conductivity is that, while $G_{yx}$  independent of $x$ for $x\gg L$, $G_{yx}$ oscillates with $x$ for $x\ll 0$. This is because, for $x\gg L$,  $\psi^{\dag}\psi=|t_k|^2$ and for $x\ll 0$,  $\psi^{\dag}\psi=(1+|r_k|^2+r_ke^{-2ik_xx}+r^*_ke^{2ik_xx})$. From eq.~\eqref{eq:Iy}, it can be seen that the transverse conductivity depends on $\psi^{\dg}\psi$, which in turn depends on $x$ for $x<0$. In Fig.~\ref{fig:GyxLL}, $G_{yx}$ is plotted versus $x$ for $\th=\pi/4$ and $\phi=0$ keeping other parameters the same. It can be seen that the transverse conductivity is not periodic in $x$ in the region $x<0$ even though it oscillates. These oscillations of transverse conductivity are rooted in density modulations  that are akin to Friedel oscillations.

We now study the dependence of $G_{xx}$ and $G_{yx}(x\gg L)$ on the length of the spin-orbit coupled region $L$. In Fig.~\ref{fig:GvsL}, we plot the two conductivities versus the length $L$ for $\th=\pi/4$ and $\phi=0$ keeping other parameters same as before. We find that the both the conductivities oscillate with $L$ though not perfectly in a periodic fashion. Such an oscillation of conductivities is reminiscent of Fabry-P\'erot type interference~\cite{liang2001,ofek2010,soori12,soori17,nehra19,soori19,suri21,soori2021,soori22car,soori2022nh}. In similar one dimensional systems, the peaks in conductance versus $L$ graph are uniformly spaced. Because, the Fabry-P\'erot resonance condition is $k \delta L=\pi$, where $\delta L$ is the difference between two consecutive peak positions. However, the conductivities in Fig.~\ref{fig:GvsL} are calculated by taking integral of angles of incidence in the range $(-\pi/2,\pi/2)$ [see eq.~\eqref{eq:Gx} and eq.~\eqref{eq:Gy}]. For different values of angle of incidence $\chi$, $\delta L$ is different. This is the reason why the peaks in Fig.~\ref{fig:GvsL} are not uniformly spaced. For the parameters chosen, the wavenumbers $k_{x,j}$ in the SOC region for normal incidence at $E=0$ are $\pm 1/a, \pm 2/a$. This means, $\delta L\sim 1.57 a$. On inspection, the average $\delta L$ in Fig.~\ref{fig:GvsL} is $1.4a$ for $G_{yx}$ and $1.6a$ for $G_{xx}$, which approximately agree with $\delta L$ from Fabry-P\'erot resonance condition for normal incidence. 

\begin{figure}[htb]
 \includegraphics[width=4cm]{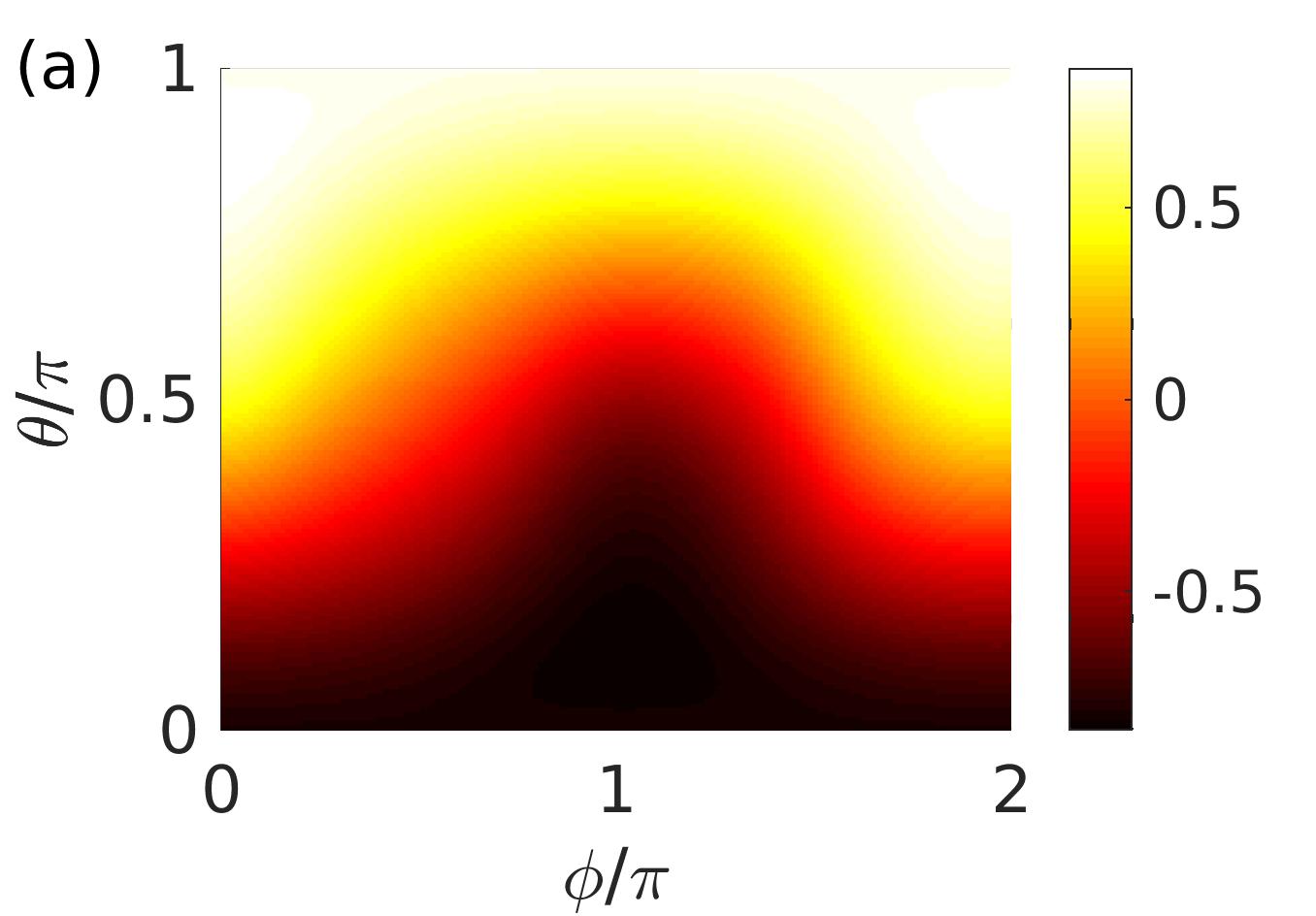}
 \includegraphics[width=4cm]{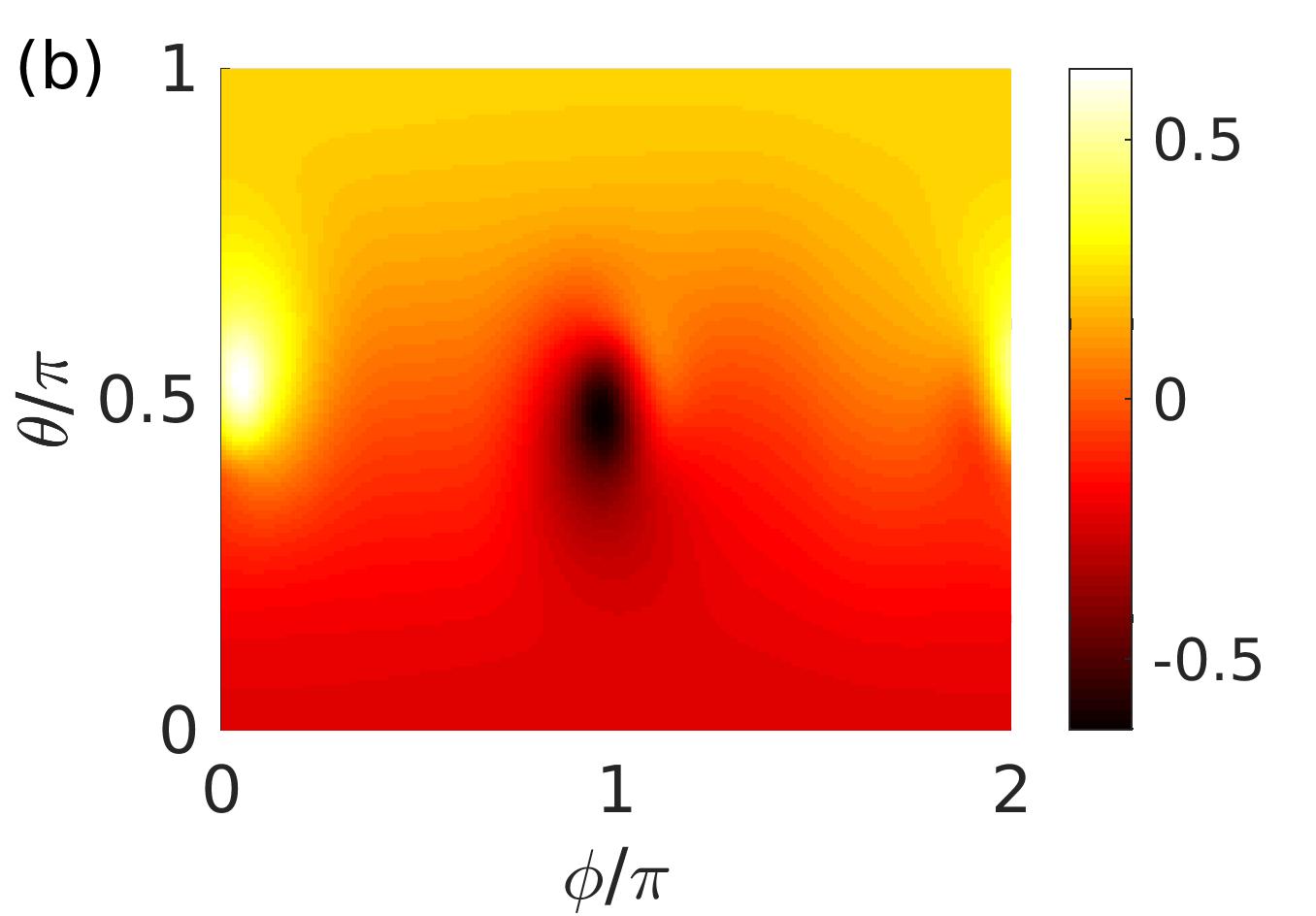}
 \includegraphics[width=8.3cm]{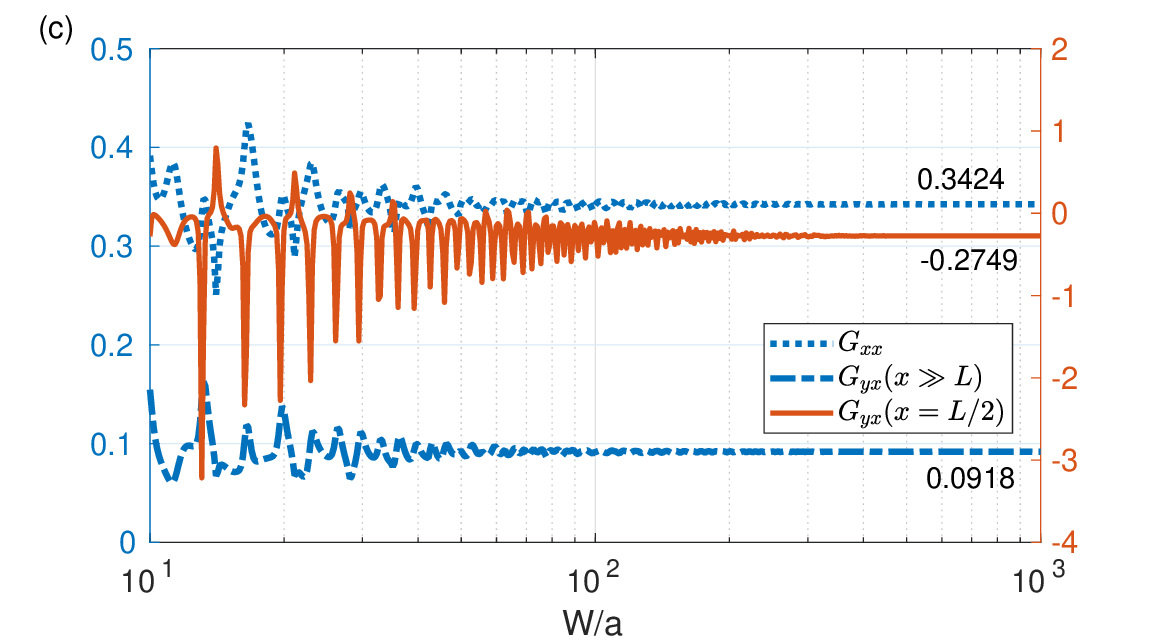}
 \caption{Transverse conductivity at $x=L/2$ for (a)$W=10a$, (b) $W=100a$. (c) Conductivities in units of $e^2/(ha)$ as functions of the width $W$ for $\th=\pi/4$, $\phi=0$. The saturation value for large $W$ for each curve is indicated in the plot. The $y$-coordinates for $G_{xx}$ and $G_{yx}(x\gg L)$ are shown on the left and the $y$-coordinate for $G_{yx}(x=L/2)$ is shown on the right.  Other parameters are same as in Fig.~\ref{fig:G}.}\label{fig:G-W}
\end{figure}

Next, we focus our attention on the case of finite width. The formulae in eq.~\eqref{eq:Gxx-W} and eq.~\eqref{eq:Gyx-W} are used to calculate the conductivities  numerically. In Fig.~\ref{fig:G-W}(a,b), the transverse conductivity at $x=L/2$ is plotted versus $(\th,\phi)$ for $W=10a, 100a$ keeping other parameters same as earlier.  For $W=1000a$, the plot of transverse conductivity versus $(\th,\phi)$ (not shown) looks similar to Fig.~\ref{fig:G}~(d) which corresponds to the transverse conductivity in the limit of infinite width $W$, whereas for $W=10a, 100a$, the  transverse conductivity plots do not resemble the one for infinite width. In Fig.~\ref{fig:G-W}(c), the conductivities $G_{xx}$, $G_{yx}(x\gg L)$ and $G_{yx}(x=L/2)$ are plotted as functions of the width $W$ for the choice $\th=\pi/4$ and  $\phi=0$. We find that the saturation values of the conductivities in the limit of large $W$ ($W\gtrsim1000$) agree with the respective conductivities calculated for the case of $W\to\infty$. 

\begin{figure*}[htb]
 \includegraphics[width=5.5cm]{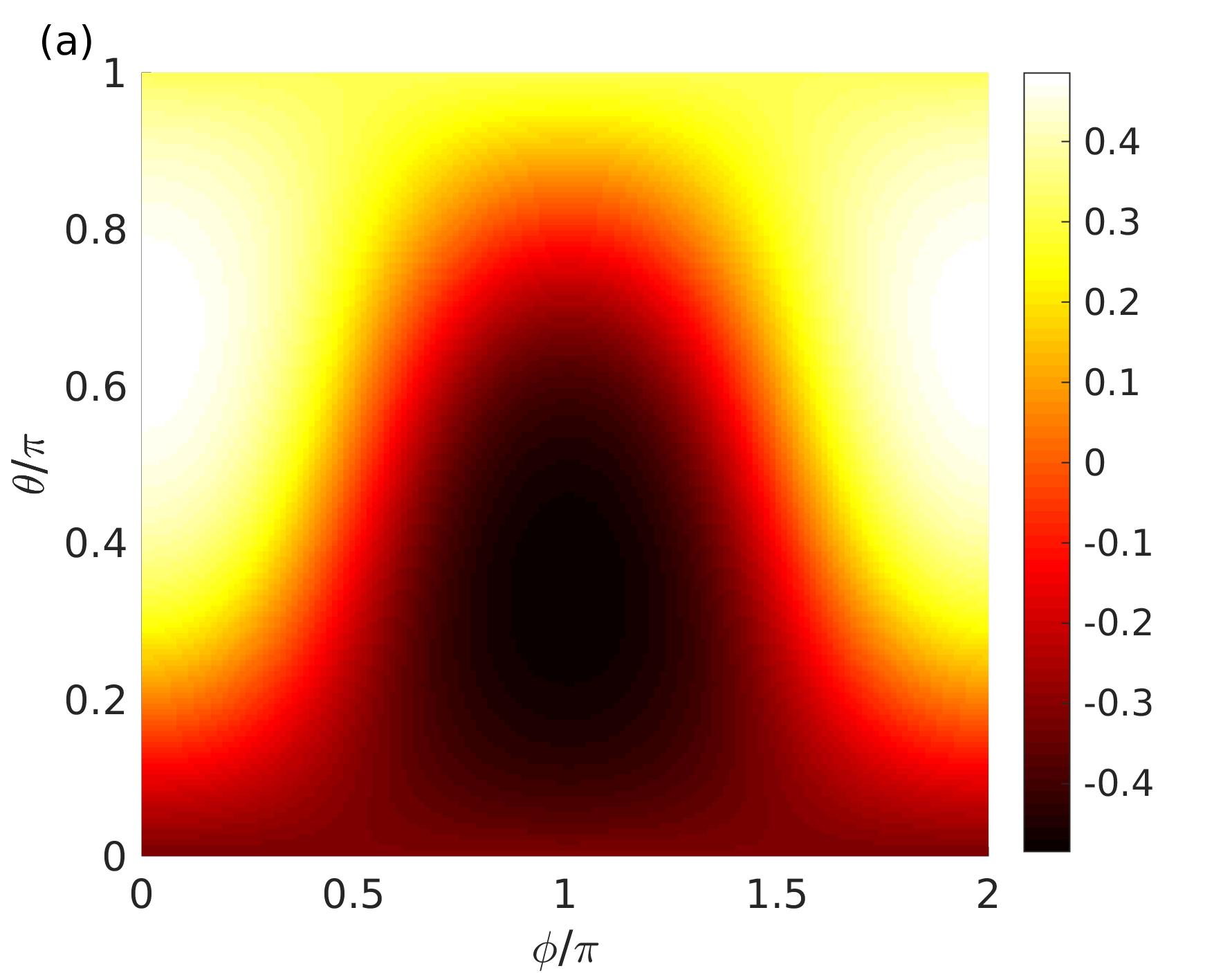} 
 \includegraphics[width=5.5cm]{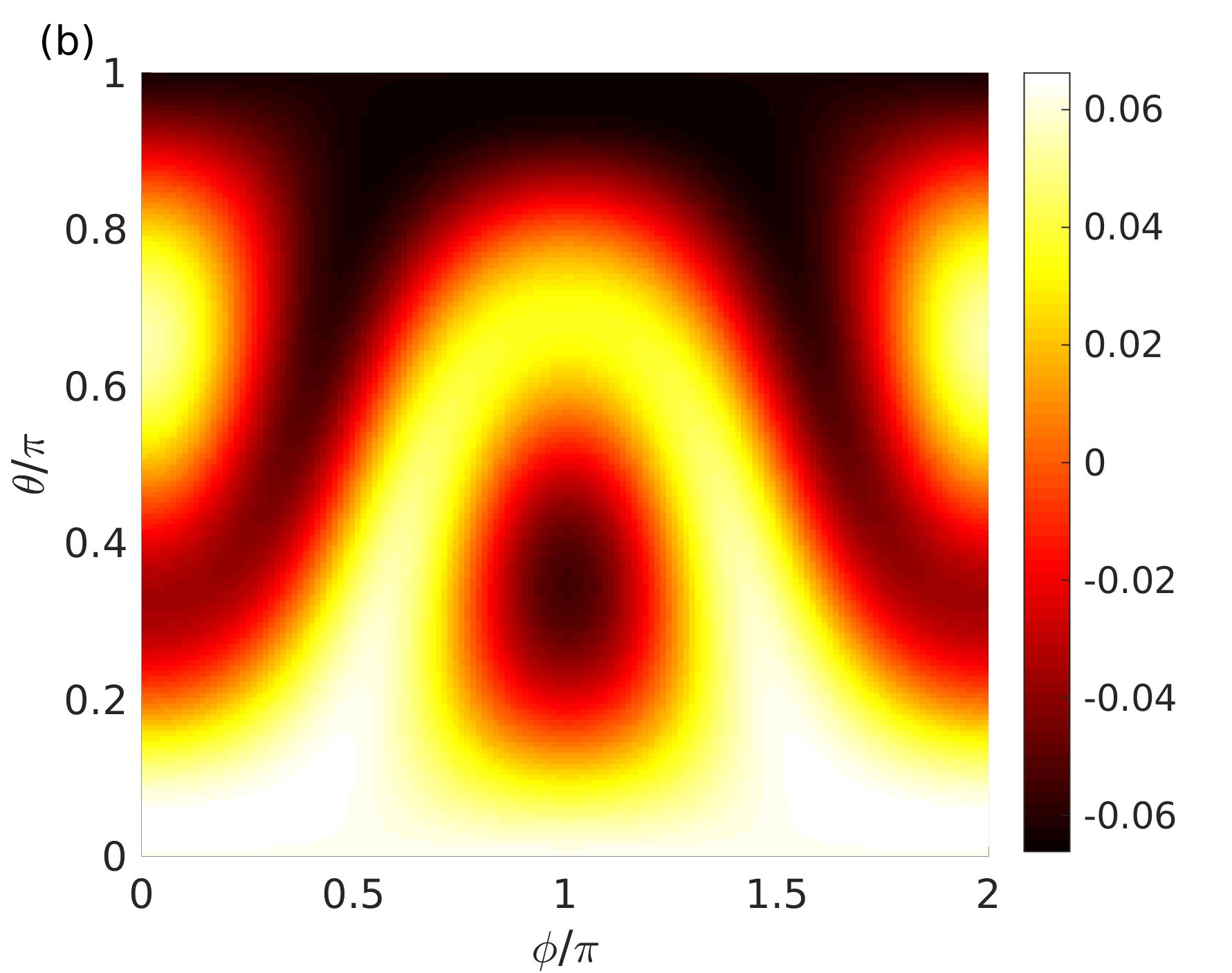}
 \includegraphics[width=5.5cm]{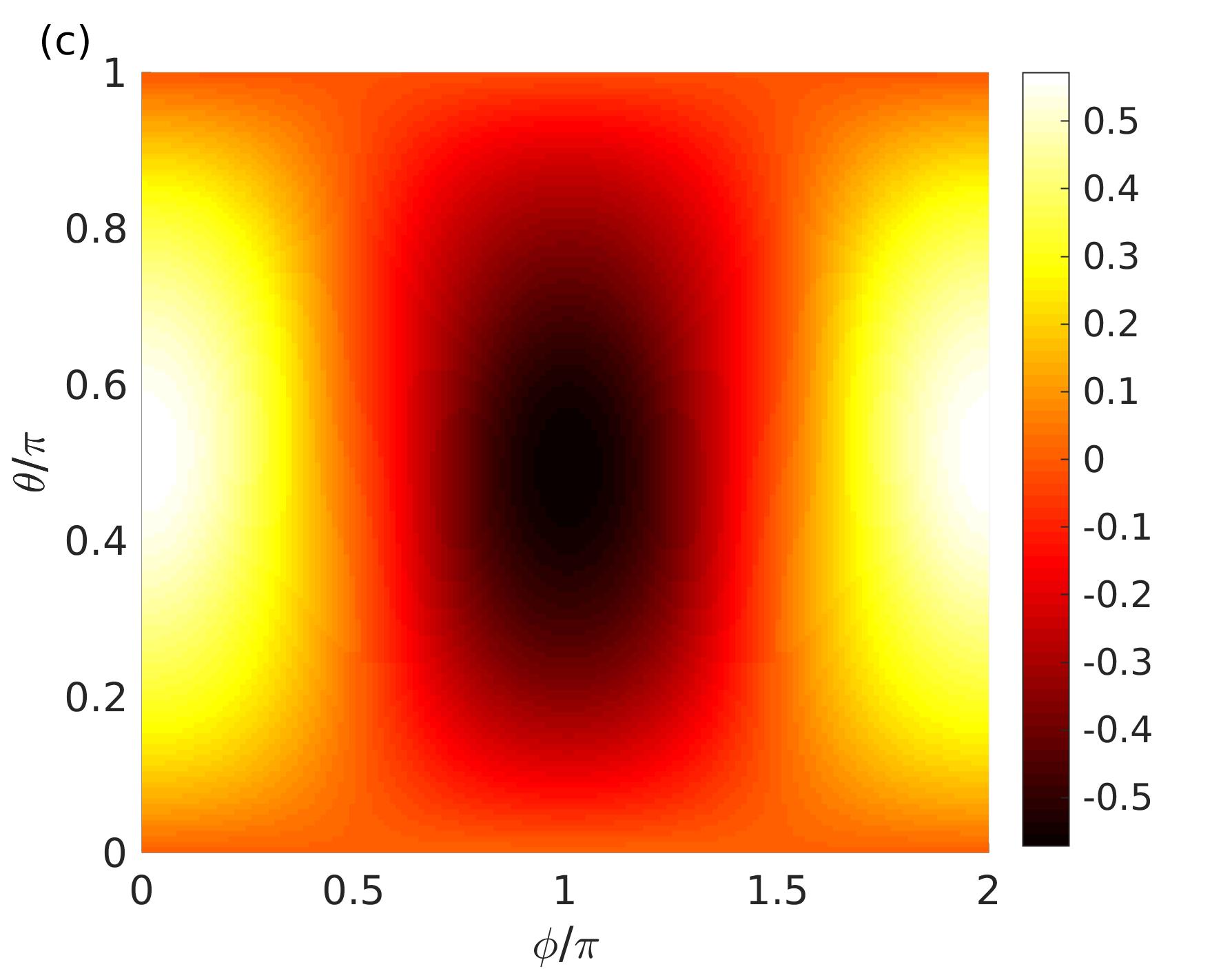}
 \caption{Differential transverse conductivity $G_{yx}(x_0)$ in units of $e^2/(ha)$ versus $(\th,\phi)$ at zero bias evaluated at (a) $x=-10a$. (b) $x=0^-$ and (c) $x=L/2$ for  $W=10a$, and other parameters same as in Fig.~\ref{fig:G}}\label{fig:Gyx-cutoff}
\end{figure*}

When the junction between spin-orbit coupled region and the ferromagnet at $x=L$ is cut-off by taking $c(L)=0$ in the boundary condition given by eq.~\eqref{eq:BC}, the longitudinal conductivity is zero. In this limit, we calculate the transverse conductivities in the region $x<L$. Zero bias transverse differential conductivities at $x=-10a,~0^-,~L/2$ are plotted as functions of $(\theta,\phi)$ in Fig.~\ref{fig:Gyx-cutoff}(a,b,c). It is interesting to see that the transverse conductivities can be nonzero even when the longitudinal conductivity is zero. This is due to a combination  of spin-polarization of incident electrons and spin orbit coupling in the central region. 

\section{Discussion}
Transverse currents are due to spin polarised electrons injected into the spin orbit coupled region. So, in the case when the region to the right of SOC region is a normal metal with no spin polarization instead of a ferromagnet, transverse currents are expected. In fact, even in the extreme case that there is nothing on the right side of SOC region (which is same as cutting of the right junction), transverse currents will appear. 

The spin Hall effect involves the generation of a transverse spin current and spin accumulation at the edges in the transverse direction as a result of a charge current flowing in the longitudinal direction~\cite{murakami03,sinova04,shermp}. On contrary, in inverse spin Hall effect, a spin current will result in transverse charge current~\cite{shermp}. The transverse current we found in this study is similar in spirit to inverse spin Hall effect. In such effects, spin Hall angle - the ratio of dimensionless charge current to dimensionless spin current determines the extent of spin to charge conversion. 
The longitudinal current is spin current, since the leads are fully spin polarized. So, the spin Hall angle is the ratio of transverse to the longitudinal conductivities. From Fig.~\ref{fig:G}, it can be seen that the spin Hall angle depends on the location at which the transverse conductivity is calculated. Spin Hall angle  reaches a maximum value of around  $\sim 0.3$ as can be seen from Fig.~\ref{fig:G}. For the case of the right junction being cut-off wherein the spin current is zero, the spin Hall angle shoots up to infinity.

An earlier work~\cite{sarkar2020} finds zero transverse currents. The main difference between their model and ours is that in their model, the spin-orbit coupled region is one-dimensional whereas in our work the spin-orbit coupled region is two-dimensional. Having two-dimensional spin-orbit coupled region is essential for getting nonzero transverse conductivity. In the limit of small $W$ for which there is only one transport channel ($N_y=0$), we find that the transverse conductivity is zero in all regions. 

Spin-orbit coupling can be induced in graphene by proximitizing it with transition metal dichalcogenides~\cite{belayadi23,belayadi23b}. Inducing ferromagnetism in graphene has also been discussed in literature~\cite{haugen,soori18}. While the strength of spin-orbit coupling in Datta-Das transistor is varied to control the conductance, gate voltage can be used to control the conductances in graphene heterostructures~\cite{belayadi23}. It will be an interesting  future direction to explore transport in spin transistors designed on graphene due to its high electron mobility and the Dirac nature. Our calculations can be repeated using the Dirac Hamiltonian for graphene with terms added for spin-orbit coupling and ferromagnetism to model Datta-Das transistor on graphene. 

\section{Summary and Conclusion}\label{sec:concl}
We have theoretically studied longitudinal and transverse conductivities in a spin transistor consisting of ferromagnets and a two-dimensional electron gas with Rashba SOC using Landauer-B\"uttiker approach. We have shown that the transverse conductivity can be non-zero and depends on the location where it is calculated, in addition to the direction of spin polarization of the ferromagnets. When the spin polarization component along $x$ in the ferromagnets is zero, the transverse conductivity is expected to be zero since the SOC is of Rashba type. However, we find that the transverse conductivity is nonzero in and around the spin-orbit coupled region. This is because of the presence of decaying modes in the ferromagnet pointing opposite to the spin polarization direction near the junctions. We explain this qualitatively by showing that $\la\si_x \ra$ near the junctions as a function of angle of incidence $\chi$ is not an odd function. The transverse conductivity oscillates with the location in the source electrode, since the probability density oscillates in the source region. The conductivities oscillate as a function of the length of the spin-orbit coupled region due to Fabry-P\'erot interference. However, the oscillation is not perfectly periodic, since the conductivities have contributions from all possible angles of incidence.  We also study the case of the system with a finite width. The conductivities oscillate as a function of the width for small and intermediate widths and for large width, saturate to the respective conductivities calculated for infinite width limit. Finally, we find that the transverse conductivity can be nonzero even in the limit when the longitudinal conductivity is zero.

 Our findings can be tested in devices wherein the components that make the transistor are ballistic.  In early experimental realizations of the spin transistors~\cite{koo09,Wunderlich2009}, the spin-orbit coupled regions were diffusive. The versions of spin transistors realized later~\cite{chuang2015,Choi2015} employed materials wherein the transport is ballistic. With the development of high quality ballistic devices in recent times, we envisage that our predictions can be put to test.

\acknowledgements
We thank DST-INSPIRE Faculty Award (Faculty Reg. No.~:~IFA17-PH190) for financial support. We thank Dhavala Suri, Amnon Aharony and Kingshuk Sarkar for stimulating discussions. 
\bibliography{ref_phe}

\end{document}